# NEGATIVE MOMENTUM COMPACTION AT KEKB

H. Ikeda, J. W. Flanagan, H. Fukuma, S. Hiramatsu, T. Ieiri, H. Koiso, T. Mimashi and T. Mitsuhashi, KEK, High Energy Accelerator Research Organization, Ibaraki 305-0801, Japan

*Abstract*

KEKB [1] is a high luminosity $e^+e^-$ collider for studying B mesons and has achieved the design luminosity of $10^{34} cm^{-2}s^{-1}$ in 2003. In order to get higher luminosity, we tested negative momentum compaction optics in the summer of 2003. We measured the bunch length using three methods at 0.7mA to 1.17mA bunch current and confirmed the length was shortened with the negative momentum compaction optics.

## INTRODUCTION

KEKB is an $e^+e^-$ collider for the B meson physics consisting of an electron ring of 8 GeV (HER) and a positron ring of 3.5 GeV (LER) as shown in Figure1. Figure 2 shows the trend of peak luminosity in the world. KEKB achieved 10.57 /nb/s on May 13th, 2003. Figure 3 shows the integrated luminosity of KEKB. We logged more than 150 /fb luminosity by 2003 summer.

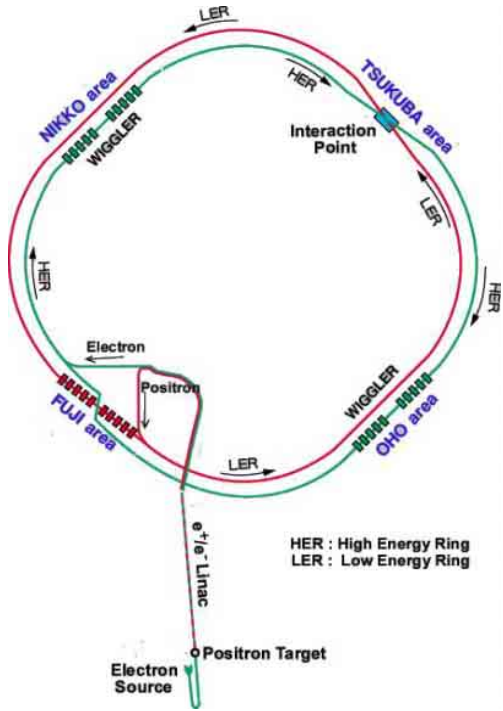

Figure 1: KEKB Accelerator.

Table 1 compares the major machine parameters between design value and the value when the best peak luminosity was achieved in 2003 summer. In order to get higher luminosity, $\beta_y^*$ was squeezed down below the design value, resulting a vertical beam-beam parameter $\xi_y$ larger than the design value. In several machine studies on higher luminosity, shortening the bunch length with a negative momentum compaction factor ($\alpha$) is one of the key issues. In the usual operation of KEKB, the bunch length is longer than design value and it may cause a luminosity reduction due to the hourglass effect. We tried negative $\alpha$ optics in order to shorten the bunch length [2]. The machine study was done on June 26th, 2003.

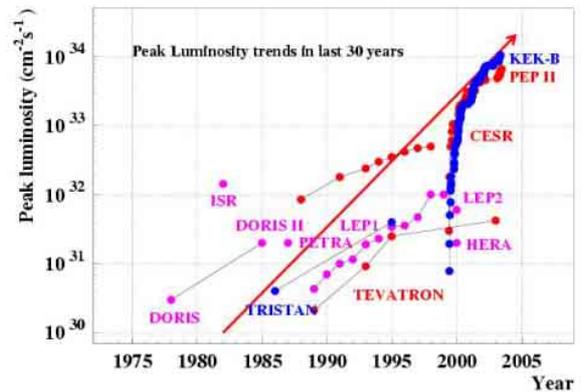

Figure 2: Trend of peak luminosity in the world over the past 30 years.

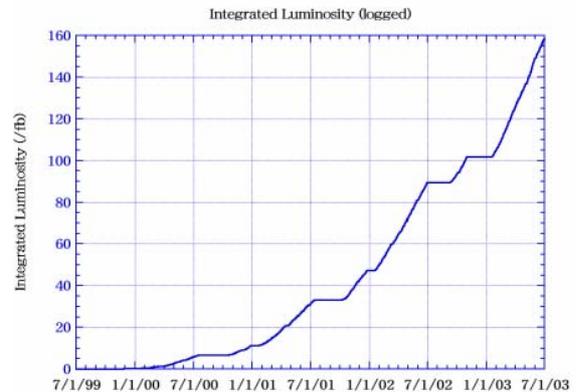

Figure 3: Integrated luminosity of KEKB.

## KEKB LATTICE

In order to achieve a large dynamic aperture, a non-interleaved chromaticity correction scheme [3, 4] has been adopted at KEKB. The $2.5\pi$ cell is created by combining five $\pi/2$ cells as shown in Figure 4. In this structure, the bending magnets are arranged to form two dispersion bumps and keep dispersion low at the dipole magnets. Pairs of sextupole magnets connected by the pseudo -I' transfer matrix are placed in the arc of the ring. By this scheme, the horizontal emittance $\varepsilon_x$ and the momentum compaction factor $\alpha$ can be adjusted independently.

Table 1: Machine Parameters

| Date | 5/13/2003 | | Design | | unit |
|---|---|---|---|---|---|
| Ring | LER | HER | LER | HER | |
| Current | 1.38 | 1.05 | 2.6 | 1.1 | A |
| Bunches | 1265 | | 5000 | | |
| Bunch Current | 1.09 | 0.83 | 0.52 | 0.22 | mA |
| Bunch Spacing | 1.8 or 2.4 | | 0.6 | | m |
| Emittance $\varepsilon_x$ | 18 | 24 | 18 | 18 | nm |
| $\varepsilon_y/\varepsilon_x$ | 4.7 | 2.9 | 2 | 2 | % |
| $\beta_x^*$ | 59 | 58 | 33 | 33 | cm |
| $\beta_y^*$ | 0.58 | 0.70 | 1.0 | 1.0 | cm |
| Hor. Beam Size @IP | 103 | 118 | 80 | 80 | μm |
| Ver. Beam Size @IP | 2.2 | 2.2 | 1.9 | 1.9 | μm |
| Beam-beam Parameter $\xi_x$ | .093 | .068 | .039 | .039 | |
| Beam-beam Parameter $\xi_y$ | .065 | .051 | .052 | .052 | |
| Luminosity | 10.57 | | 10 | | /nb/s |
| $\int Lum/day$ | 579 | | 600 | | /pb |
| $\int Lum/7days$ | 3876 | | | | /pb |
| $\int Lum/30days$ | 12809 | | | | /pb |

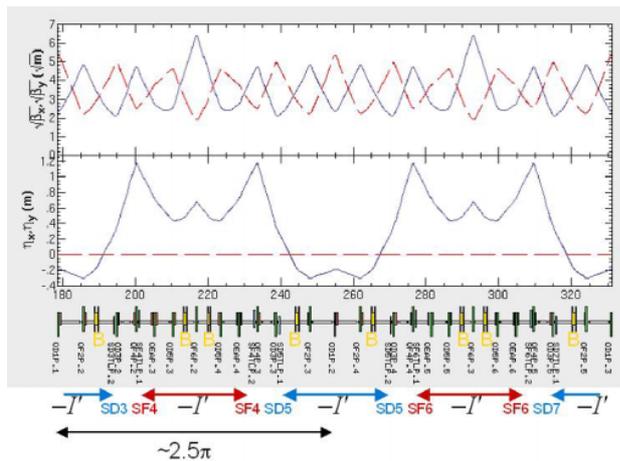

Figure 4: 2.5 π cell of KEKB lattice.

We tried the negative α optics where the absolute value of α is almost same to the usual one. Both the HER and LER optics are shown in Figure 5. Table 2 shows the machine parameters for the study. Measured synchrotron tunes were consistent with set values. Results of beam phase and bunch length measurement are shown in the following sections.

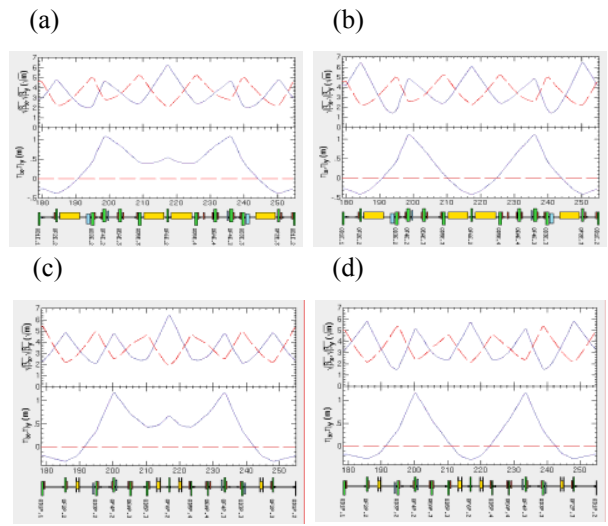

Figure 5: KEKB Lattice for LER (a, b) and HER (c, d). (a) and (c) are those of positive momentum compaction, and (b) and (d) are those of negative momentum compaction cells.

Table 2: Machine parameters for study

|  | LER |  | HER |  |
|---|---|---|---|---|
| Hor. Emittance $\varepsilon_x$(nm) | 18.8 | 18.1 | 24.1 | 24.0 |
| Compaction Factor $\alpha$ | 3.41E-4 | -3.41E-4 | 3.38E-4 | -3.45E-4 |
| Bunch Length $\sigma_z$(mm) | 4.75 @8MV | 4.58 @8MV | 5.22 @13MV | 5.26 @13MV |
| Synchrotron Tune $\nu_{s\,set}$ ($\nu_{s\,measured}$) | 0.0249 | 0.0248 (0.0247) | 0.0208 | 0.0209 (0.0206) |
| Betatron Tune $\nu_x/\nu_y$ | 45.508/ 43.543 | 47.519/ 43.560 | 44.509/ 41.587 | 44.57/ 41.60 |

## BEAM PHASE MEASUREMENT

Switching from a positive to a negative α lattice changes the synchronous phase φs as shown in Figure 6 [2]. The shift in the synchronous phase due to the parasitic loss Δφs is given by

$$V_c \sin(\phi_{s0} + \Delta\phi_s) = \frac{U_0}{e} + k(\sigma)T_0 I_b, \quad (1)$$

where $V_s$ is the accelerating cavity voltage, $\phi_{s0}$ is the synchronous phase of a zero-current limit, $U_0$ is the radiation loss per turn, k (σ) is the loss factor, and $T_0$ is the revolution period. The relative beam phase was measured as a function of beam current for negative and positive α optics as shown in Figure 7. Assuming that the RF phase is constant, we can estimate the $\phi_{s0}$ from an extrapolated phase to zero current and the difference between negative and positive α results Δψ. The measured $\phi_{s0}$ was 10.3 degree at the LER and 16.7 degree at the HER. It was consistent with expectation for both rings.

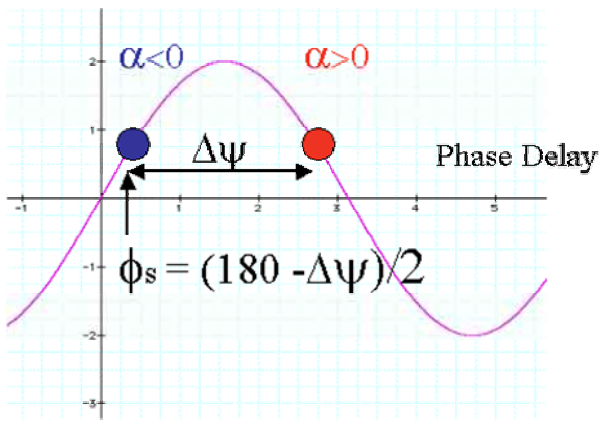

Figure 6: Synchrotron phase changes with momentum compaction sign.

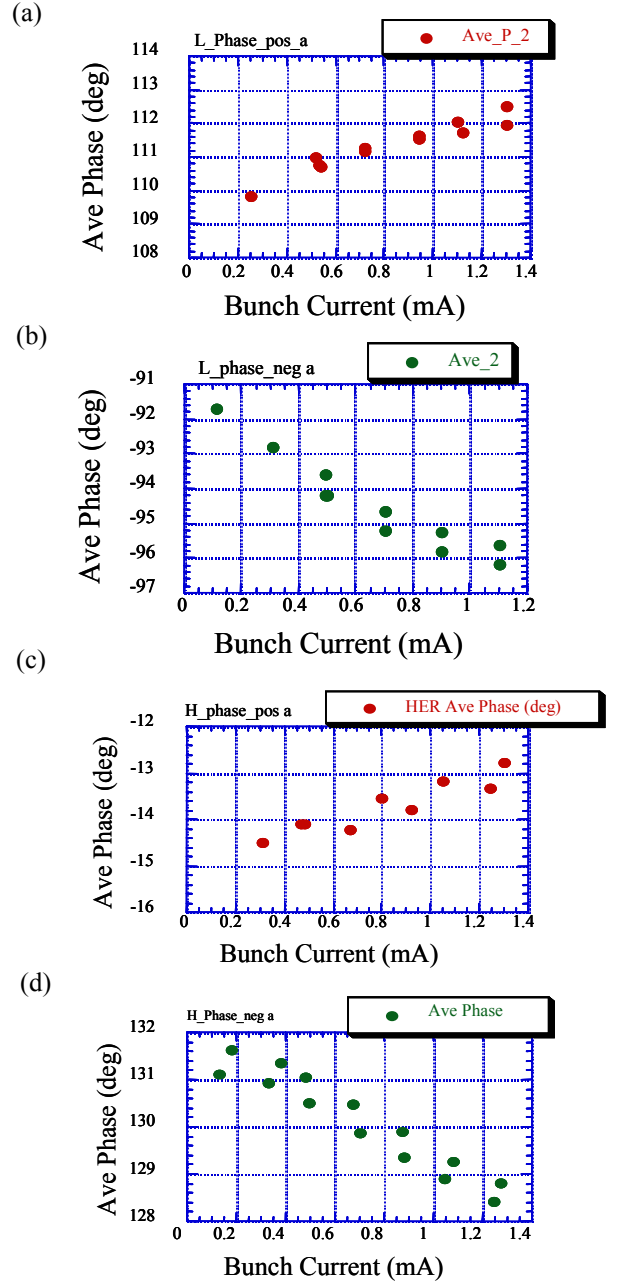

Figure 7: Measured beam phase of LER (a, b) and HER (c, d). (a) and (c) are positive α optics and (b) and (d) are negative α optics.

## BUNCH LENGTH MEASUREMENT

The bunch length was measured for both negative and positive α optics. We have three methods to measure the bunch length for KEKB [5]. First is the RMS bunch length monitor, second is the RF wave-guide system and third is the streak camera. All three methods were used for this study to check the consistency.

### RMS Bunch Length Monitor

The bunch length is evaluated by detecting two frequency components of the bunch spectrum as

$$\sigma_l = c\sqrt{\frac{2}{(\omega_2^2 - \omega_1^2)}\ln(\frac{V_1}{V_2})}, \quad (2)$$

where $V_1$ and $V_2$ are the spectrum amplitudes of the bunch signal at frequencies of $\omega_1$ (=2π1.02GHz) and $\omega_1$ (=2π2.54 GHz).

In this monitor, the bunch signal is picked up by a button electrode installed on a beam pipe and the bunch length is calculated by an analog calculator unit for the real time measurement. Figure 8 shows the relation between the amplitude ratio and bunch length. The resolution of the measurement is estimated to be about 0.2mm at 4mm bunch length.

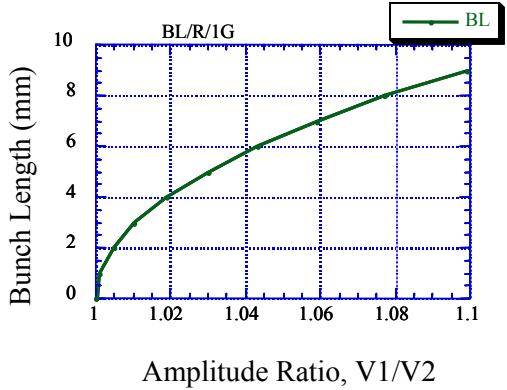

Figure 8: Bunch length measurement based on beam spectrum.

### RF Wave-Guide System

The bunch length can be evaluated from a beam spectrum measurement of the beam-induced field in an RF cavity. In order to detect the field, a wideband pickup is mounted on the wave-guide of the RF system (Figure 9 (a)). Since the wavelengths of the field components above 5GHz are much smaller than the size of wave guide, almost all components pass through the wave-guide. The bunch length is estimated by fitting the spectrum envelope using a Gaussian with the fitting parameters of the bunch length $\sigma_l$ and the normalization factor a as shown in Figure 9 (b). The fitting function is

$$F(\omega) = \frac{a\sigma_l}{\sqrt{2\pi}}e^{-\frac{\omega^2\sigma_l^2}{2}}\sum_{n=-\infty}^{\infty}e^{j\omega n\Delta T}, \quad (3)$$

assuming that the bunched beam passes through the RF cavity every ΔT sec.

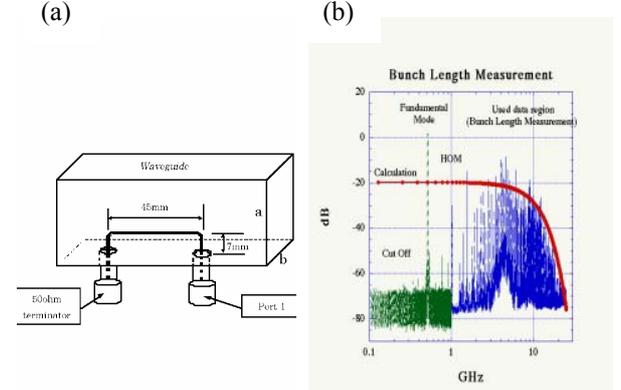

Figure 9: (a) A pickup mounted on the wall of RF wave-guide system. (b) Beam spectrum measured through the wave-guide system.

### Streak Camera

The direct observation of the bunch length is performed using a streak camera with reflected optics to increase the light intensity of the synchrotron radiation of the individual beam bunch produced in the weak bending magnet installed in the KEKB ring. The bunch-by-bunch shapes were observed as shown in Figure 10. In the case of negative α optics, the bunch shape is asymmetrical. It may be a consequence of potential well distortion effects.

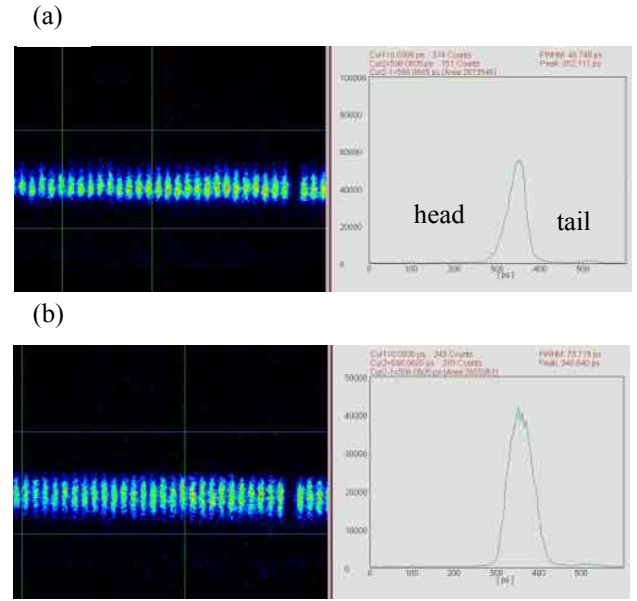

Figure 10: Bunch shape measured by streak camera at (a) negative and (b) positive α optics. Time goes left to right.

### Results

Figure 11 shows the results of three methods for bunch length measurement in the LER. Solid marks show the negative α optics and open marks show the positive α optics. All of three methods for the negative α case show a shorter bunch length than that of positive α optics. The shortening of the bunch length is enhanced as the bunch current increases although the calculation shown in Table 2 predicts a small effect on the bunch length in the zero-current limit. The bunch length of the HER was measured only by the RMS bunch length monitor (Figure 12) and it was also shorter at negative α optics.

## SUMMARY

The flexible KEKB lattice enables us to adjust the horizontal emittance and momentum compaction factor independently. Lattice setting of negative α optics was successfully done. The synchrotron tune is consistent with the calculated value and the beam phases were changed between negative and positive α optics as expected. We confirmed that bunch lengthening was reduced at negative α optics using the three methods. The bunch length shortening using the negative α lattice has the potential to increase the KEKB luminosity further.

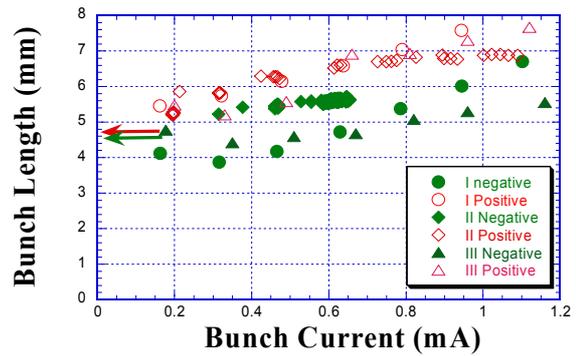

Figure 11: Bunch length measurement for the LER at KEKB using three methods. (I: RMS bunch length monitor, II: RF wave-guide system, and III: streak camera result.) Arrowheads show the expected natural bunch length.

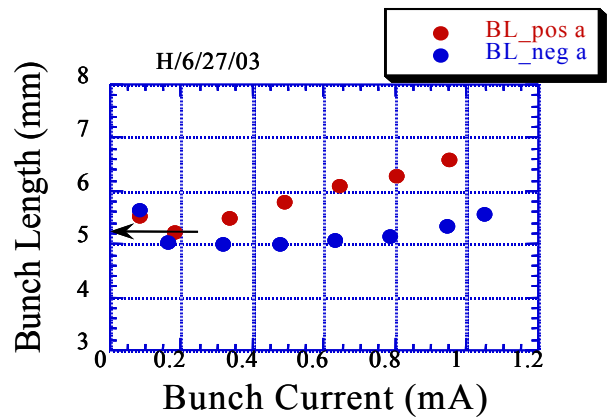

Figure 12: Bunch length measurement for the HER at KEKB using the RMS bunch length monitor.